\documentclass[twocolumn,showpacs,preprintnumbers,amsmath,amssymb,prb]{revtex4}

\usepackage{graphicx,here}
\usepackage{dcolumn}
\usepackage{bm}
\usepackage{enumerate}

\begin{document}

\preprint{APS/123-QED}

\title{Magnons and electromagnons in a spin-lattice-coupled frustrated magnet CuFeO$_2$\\
as seen via inelastic neutron scattering}

\author{Taro Nakajima}
\email{E-mail address: nakajima@nsmsmac4.ph.kagu.tus.ac.jp}
\author{Azusa Suno}
\altaffiliation{present address: Insititute of Physics, University of Tsukuba, Tennodai, Tsukuba 305-8571,Japan}
\author{Setsuo Mitsuda}
\affiliation{Department of Physics, Faculty of Science, Tokyo University of Science, Tokyo 162-8601, Japan}%
\author{Noriki Terada}
\affiliation{National Institute for Materials Science, Tsukuba, Ibaraki 305-0044, Japan}
\author{Shojiro Kimura}
\affiliation{Institute for Materials Research, Tohoku University, Sendai 980-8577, Japan.}
\author{Koji Kaneko}
\author{Hiroki Yamauchi}
\affiliation{Quantum Beam Science Directorate, Japan Atomic Energy Agency, Tokai, Ibaraki 319-1195, Japan}

\begin{abstract}
We have investigated spin-wave excitations in a four-sublattice (4SL) magnetic ground state of a frustrated magnet CuFeO$_2$, in which `electromagnon' (electric-field-active magnon) excitation has been discovered by recent terahertz time-domain spectroscopy [Seki \textit{et al.} Phys. Rev. Lett. {\bf 105} 097207 (2010)]. %
In previous study, we have identified two spin-wave branches in the 4SL phase by means of inelastic neutron scattering measurements under applied uniaxial pressure. [T. Nakajima \textit{et al}. J. Phys. Soc. Jpn. {\bf 80} 014714 (2011) ] %
In the present study, we have performed high-energy-resolution inelastic neutron scattering measurements in the 4SL phase, resolving fine structures of the lower-energy spin-wave branch near the  zone center. %
Taking account of the spin-driven lattice distortions in the 4SL phase, we have developed a model Hamiltonian to describe the spin-wave excitations. %
The determined Hamiltonian parameters have successfully reproduced the spin-wave dispersion relations and intensity maps obtained in the inelastic neutron scattering measurements. %
The results of the spin-wave analysis have also revealed physical pictures of the magnon and electromagnon modes in the 4SL phase, %
suggesting that collinear and noncollinear characters of the two spin-wave modes are the keys to understand the dynamical coupling between the spins and electric dipole moments in this system. %

\end{abstract}

\maketitle

\section{INTRODUCTION}

A triangular lattice antiferromagnet CuFeO$_2$ (CFO) has recently attracted increasing attention due to its cross-correlated phenomena arising from spin frustration. %
CFO has a delafossite structure, in which triangular lattice layers of magnetic Fe$^{3+}$ ions are separated by nonmagnetic O$^{2-}$-Cu$^{+}$-O$^{2-}$ dumbbells. %
From the electronic state of Fe$^{3+}$ ions ($S=5/2$, $L=0$) and antiferromagnetic interactions between them, one might expect a noncollinear three-sublattice magnetic ground state so called "120$^{\circ}$-structure".\cite{Collins} However, CFO exhibits, in a ground state, a collinear four-sublattice (4SL) antiferromagnetic order in which the magnetic moments of Fe$^{3+}$ ions are confined along the $c$ axis\cite{Mitsuda_1991} (see Fig. \ref{result}(a)). %
Recent theoretical studies have pointed out that in frustrated magnets, collinear magnetic orderings can be stablized by strong spin-lattice coupling.\cite{PRL2004_Penc,PhysRevLett.100.077201}%
Actually, synchrotron radiation x-ray diffraction studies by Terada \textit{et al.}\cite{Terada_CuFeO2_Xray} and Ye \textit{et al.}\cite{Ye_CuFeO2} have revealed that CFO exhibits `spin-driven' crystal structural transitions; %
while the crystal structure of CFO is a trigonal structure (space group $R\bar{3}m$) in the paramagnetic (PM) phase, it turns to be a monoclinic structure in the 4SL phase. %
This implies that the lattice degree of freedom is the key to stabilize the 4SL magnetic order. %

CFO is also known as a spin-driven magneto-electric (ME) multiferroic,\cite{Kimura_CuFeO2}  in which a screw-type magnetic order induced by nonmagnetic substitution or application of magnetic field breaks inversion symmetry of the system, and triggers ferroelectricity.\cite{Kanetsuki_JPCM,Seki_PRB_2007,Ga-induce,CFRO,SpinNoncollinearlity,CompHelicity,CFAO_Helicity} %
Furthermore, recent terahertz time-domain spectroscopy has discovered `electromagnon (electric-field-active magnon)' excitation in the \textit{nonferroelectric} 4SL phase,\cite{Seki_Electromagnon} %
indicating a dynamical coupling between spins and electric dipole moments in this system. %

To understand these various cross-correlated phenomena, it is fundamental to determine the magnetic interaction parameters of this system. %
Ye \textit{et al.} have performed inelastic neutron scattering measurements to determine these parameters from spin-wave dispersion relations in the 4SL phase.\cite{Ye_CFO_SW} 
However, in the previous work, there is some ambiguity in the identification of the spin-wave branches %
because of a problem arising from magnetic domain structures in this system. %
Specifically, in the 4SL phase, CFO has three types of magnetic domains whose magnetic propagation wave vectors are described as $(\frac{1}{4},\frac{1}{4},\frac{3}{2})$, $(-\frac{1}{2},\frac{1}{4},\frac{3}{2})$ and $(\frac{1}{4},-\frac{1}{2},\frac{3}{2})$ using a hexagonal basis. %
These wave vectors are crystallographically equivalent to each other because of the threefold rotational symmetry of the original trigonal crystal structure. %
Therefore, the magnetic excitation spectrum reported in the previous study\cite{Ye_CFO_SW}  is a mixture of spin-wave spectrums corresponding to the three different orientations of the magnetic domains. %

Quite recently, we have demonstrated that a `single-domain' 4SL phase can be realized by uniaxial pressure applied perpendicular to the $c$ axis.\cite{SingleDomainSW} %
We performed inelastic neutron scattering measurements under applied uniaxial pressure, and identified two spin-wave branches in the single-domain 4SL phase. %
We have also pointed out that the lower-energy branch splits into two branches near the  zone center. %
However, the wave-vector ($Q$) dependence of the splitting was not fully investigated. %

In the present study, we have thus performed high-energy-resolution inelastic neutron scattering measurements to resolve the fine structure of the spin-wave dispersion relations in the 4SL phase. %
Taking into account of the spin-driven lattice distortion, we have developed a model Hamiltonian to describe the observed spin-wave spectrums. %
The results of the spin-wave calculation have revealed the microscopic picture of the magnon and electromagnon modes in the 4SL phase. %

\section{EXPERIMENT AND RESULTS}

We used the single-crystal CFO sample identical to the sample used in Ref. \onlinecite{SingleDomainSW}. %
The sample has been set in a uniaxial pressure cell developed by Aso \textit{et al.}\cite{Pressurecell} %
An uniaxial pressure of 10 MPa has been applied to the $[1\bar{1}0]$ surfaces of the sample in order to produce the single-domain 4SL state in which the magnetic domains with the wave vector of $(\frac{1}{4},\frac{1}{4},\frac{3}{2})$ dominate over the others.\cite{SingleDomainSW} %

We used a cold neutron triple axis spectrometer LTAS(C2-1) installed at JRR-3 in Japan Atomic Energy Agency, Tokai, Japan. %
The energy of the scattered neutrons was fixed to $E_f = 2.5$ meV. %
A horizontal focusing analyzer was employed. %
Energy resolution at elastic position was 0.06 meV  (full width at half maximum). %
The higher-order contaminations were removed by a cooled Be-filter placed in front of the sample and a room-temperature Be-filter placed in front of the analyzer. %

\begin{figure}[t]
\begin{center}
	\includegraphics[clip,keepaspectratio,width=7.8cm]{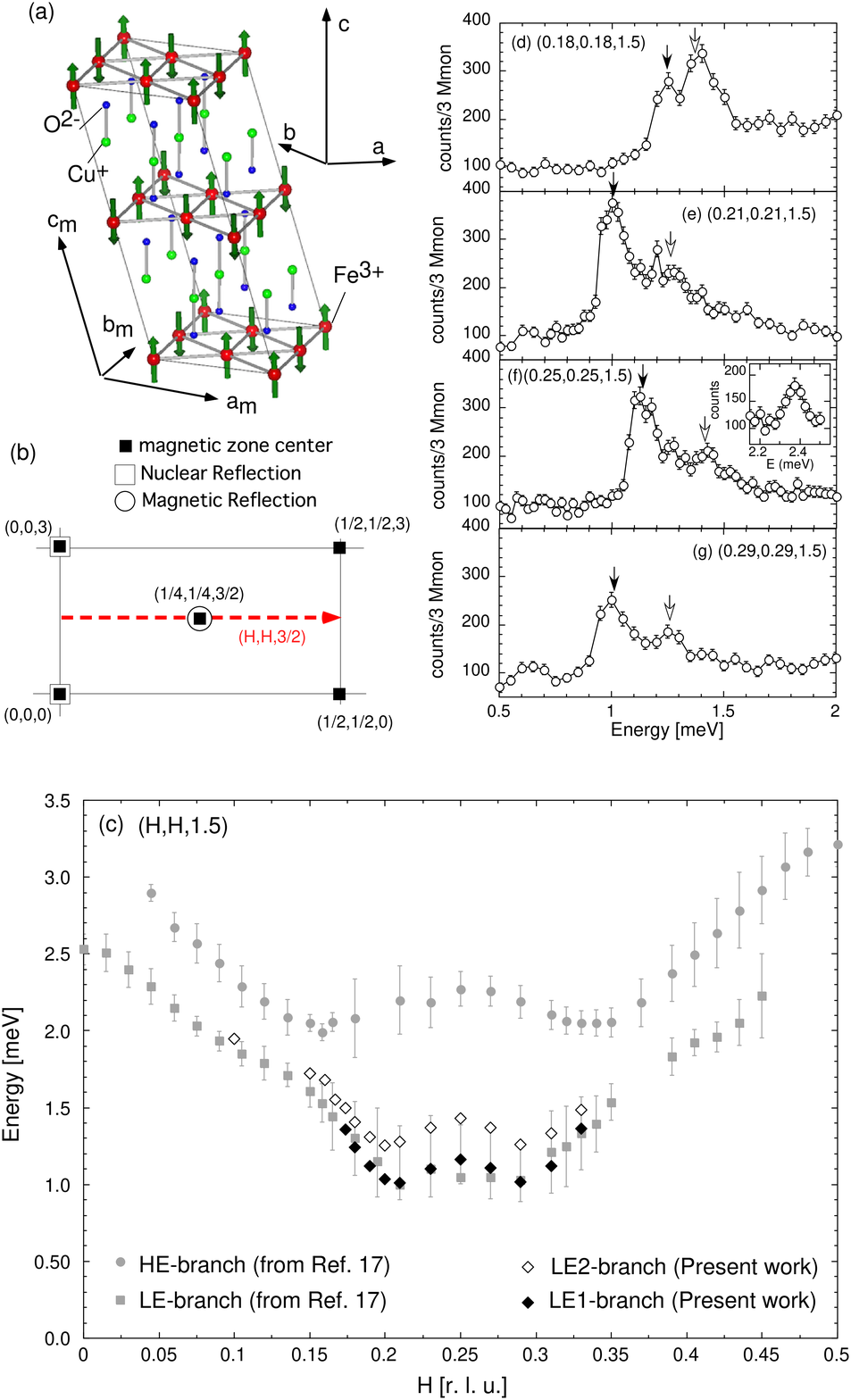}
	\caption{(Color Online) (a) A magnetic unit cell of the 4SL magnetic structure. ${\bm a}_m, {\bm b}_m$ and ${\bm c}_m$ denote the monoclinic basis. (b) The reciprocal lattice map of the $(H,H,L)$ scattering plane in CFO. (c) The peak positions of the constant-$Q$ scans combined with the results in Ref. \onlinecite{SingleDomainSW}. The horizontal bars denote  full widths at half-maximums of the scattering profiles in the previous study. [(d)-(g)] The scattering profiles of the constant-$Q$ scans at (d) (0.18,0.18,1.5), (e) (0.21,0.21,1.5), (f) (0.25,0.25,1.5) and (g) (0.29,0.29,1.5), at $T=2$ K in the 4SL phase.  }
	\label{result}
\end{center}
\end{figure}

\begin{figure}[t]
\begin{center}
	\includegraphics[clip,keepaspectratio,width=8cm]{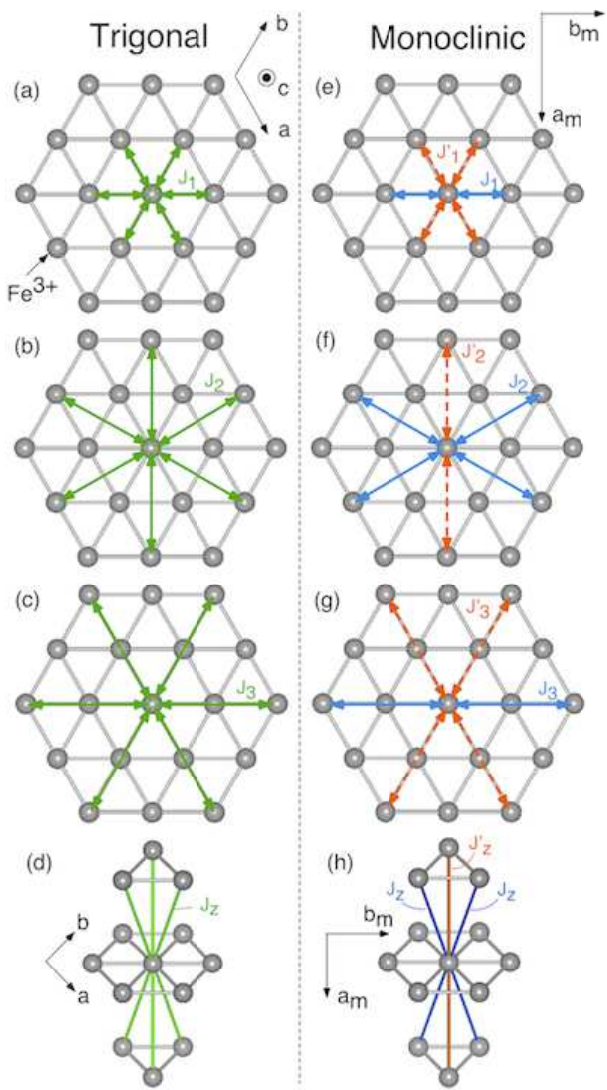}
	\caption{(Color Online) [(a)-(h)] Paths of the exchange interactions [(a)-(d)] in the trigonal symmetry and [(e)-(h)] the monoclinic symmetry of the crystal structure of CFO. }
	\label{interactions}
\end{center}
\end{figure}

As mentioned in the introduction, we have already identified two spin-wave branches in our previous work.\cite{SingleDomainSW} %
We have referred to the higher and lower energy branches as the HE- and LE-branches, respectively. %
To resolve the fine structure of these branches, we performed constant-$Q$ scans at several reciprocal lattice points at $T=2$ K in the 4SL phase, as shown in Figs. \ref{result}(d)-\ref{result}(g). %
We found that the scattering profiles apparently have two-peak structures around the energies corresponding to the LE-branch. %
On the other hand, splitting of the HE-branch was not observed within the present experimental resolution, as shown in the inset of Fig. \ref{result}(f). %
Therefore, we focused on the LE-branch, and systematically carried out constant-$Q$ scans along $(H,H,\frac{3}{2})$ line in the range of $0.15 < H < 0.35$ and $0 < E < 2.0$ meV. %
In Fig. \ref{result}(c), we have plotted the peak positions of the observed scattering profiles together with the results in Ref. \onlinecite{SingleDomainSW}, revealing the dispersion curves of the two split branches. %
Hereafter, we refer to the split branches having lower and higher energies as "LE1" and "LE2" branches, respectively. %

\section{CALCULATIONS AND DISCUSSIONS}
\subsection{Model Hamiltonian for the Spin-Wave Excitations in the 4SL Phase}

The first calculation for the spin-wave dispersion relations in the 4SL phase has been presented by Fishman.\cite{Fishman_CFO_SW} %
Although the calculation, in which three in-plane exchange interactions ($J_1, J_2$ and $J_3$), a exchange interaction between adjacent layers ($J_z$) and a uniaxial single ion anisotropy ($D$) are employed, successfully reproduces the dispersion relation of the LE branch, it does not that of the HE branch. %
Specifically, in Ref. \onlinecite{Fishman_CFO_SW}, the excitation bandwidth for the HE branch along $(H,H,3/2)$ line is calculated to be $\sim 2.5$ meV (from 2.5 to 5.0 meV), while that has been determined to be $\sim 1.2$ meV (from 2.0 to 3.2 meV) in our experiment.\cite{SingleDomainSW} %
In subsequent study, Fishman \textit{et al.} have corrected the paths of the exchange interactions between the adjacent layers, and have presented revised calculations for the spin-wave dispersion relations.\cite{Fishman_stacking} However, the dispersion relation of the HE branch has not been discussed. %
These previous calculations\cite{Fishman_CFO_SW,Fishman_stacking} are based on the assumption that all of the exchange interactions are isotropic, as shown in Figs. \ref{interactions}(a)-\ref{interactions}(d). %
In other words, effects of the spin-driven lattice distortions on the exchange interactions are not taken into account. %

\begin{figure}[t]
\begin{center}
	\includegraphics[clip,keepaspectratio,width=7cm]{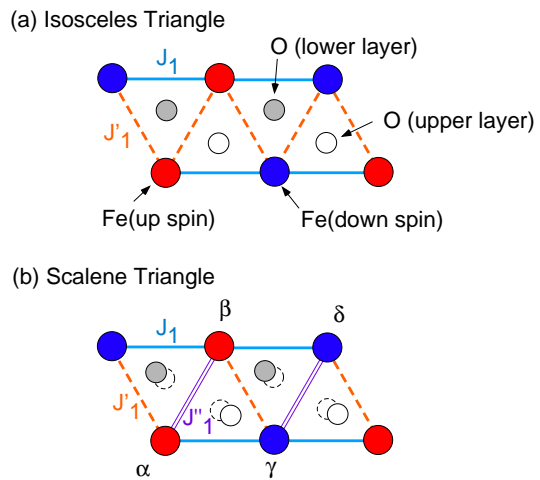}
	\caption{(Color Online) (a) Isosceles and (b) scalene triangle lattices with the 4SL magnetic order.}
	\label{scalene_triangle}
\end{center}
\end{figure}

On the other hand, Kimura \textit{et al.} have recently discussed the effects of lattice distortions on ESR signals, i.e., spin-wave energies at zone center, in the 4SL phase.\cite{Kimura_ESR_JLTP} %
They have performed multi-frequency ESR measurements in the 4SL phase, and have analyzed the results using the `scalene triangle model' proposed by the synchrotron radiation x-ray diffraction study by  Terada \textit{et al}.\cite{Terada_CuFeO2_Xray} In this model, the nearest neighbor exchange interactions, $J_1$, splits into three inequivalent interactions due to the monoclinic lattice distortion and displacements of the O$^{2-}$ ions, as shown in Fig. \ref{scalene_triangle}(b). %
These deformations are expected to enhance the antiferromagnetic exchange interactions between neighboring up and down spins, and on the contrary, to reduce those between neighboring two up (or down) spins, as was discussed in Refs. \onlinecite{Terada_CuFeO2_Xray} and \onlinecite{Terada_LatticeModulation}. %
Consequently, the splitting of $J_1$ contributes to stabilize the 4SL magnetic ordering. %

Quite recently, Kimura \textit{et al} have calculated spin-wave dispersion relations in the 4SL phase using the scalene triangle model, %
assuming that  $J_1$ splits into three inequivalent interactions, while $J_2$, $J_3$ and $J_z$ are isotropic.\cite{Kimura_unpublished} %
They have shown that the energy bandwidth of the HE branch along $(H,H,3/2)$ line is calculated to be $\sim 1.0$ meV (from 1.8 to 2.8 meV). %
In addition, their calculations have qualitatively reproduced overall features of the results of the inelastic neutron scattering measurements in the multi-domain state.\cite{Ye_CFO_SW} %
However, there still remain some discrepancies between the `shapes' of the calculated and observed dispersion curves for the single-domain 4SL phase. %

In the present study, we start from three in-plane exchange interactions $J_1, J_2$ and $J_3$ and a inter-layer exchange interaction $J_z$. %
First, we introduce effects of the trigonal to monoclinic structural transition on all of the exchange interactions. %
As shown in Figs. \ref{interactions}(e)-\ref{interactions}(h), the monoclinic symmetry of the crystal allows the exchange interaction $J_i$ $(i=1,2,3,z)$ to split into $J_i$ and $J'_i$. %
Second, we have applied the scalene triangle model to the nearest neighbor exchange interactions, and therefore $J'_1$ splits into $J'_1$ and $J''_1$, as shown in Figs. \ref{scalene_triangle}(a) and \ref{scalene_triangle}(b). %
Although the scalene triangle distortion might affect the other distant interactions, we have assumed that it affects only the nearest neighbor interactions in the present analysis. %

As for the single ion anisotropy, Fishman \textit{et al.} have employed only a uniaxial anisotropy,  $D$.\cite{Fishman_CFO_SW,Ye_CFO_SW,Fishman_stacking} %
On the other hand, Kimura \textit{et al.} have argued that the in-plane anisotropy $E$ is necessary to explain the results of their ESR measurements.\cite{Kimura_ESR_JLTP} %
We have thus employed $D$ and $E$ terms in the present analysis. %
Therefore, we have written a Hamiltonian for the 4SL phase as 
\begin{eqnarray}
\mathcal{H} =&& -\frac{1}{2}\sum_{i,j}\sum_{n=1}^{3} J_n \mbox{\boldmath $S$}_i\cdot\mbox{\boldmath $S$}_j%
-\frac{1}{2}\sum_{i,j} \sum_{n=1}^{3}J'_n \mbox{\boldmath $S$}_i\cdot\mbox{\boldmath $S$}_j\nonumber\\%
&&-\frac{1}{2}\sum_{i,j} J''_1 \mbox{\boldmath $S$}_i\cdot\mbox{\boldmath $S$}_j\nonumber\\
&& -\frac{1}{2}\sum_{i,j}J_z \mbox{\boldmath $S$}_i\cdot\mbox{\boldmath $S$}_j%
-\frac{1}{2}\sum_{i,j} J'_z \mbox{\boldmath $S$}_i\cdot\mbox{\boldmath $S$}_j\nonumber\\%
&&-\sum_{i}D(\mbox{\boldmath $S$}_i^z)^2-\sum_{i}E[(\mbox{\boldmath $S$}_i^x)^2-(\mbox{\boldmath $S$}_i^y)^2],%
\label{Hamiltonian}
\end{eqnarray}
where the $x, y$ and $z$ axes are defined to be parallel to $[110]$, $[\bar{1}10]$ and $[001]$ directions of the crystal, respectively. %

\subsection{Calculation for the Spin-Wave Spectrum}

We have calculated the spin-wave energies in the 4SL phase, %
applying a Holstein-Primakoff 1/S expansion about the classical limit to the Hamiltonian of Eq. (\ref{Hamiltonian}). %
Along the previous work by Fishman,\cite{Fishman_CFO_SW} we refer to the four magnetic sublattices as $\alpha, \beta, \gamma$ and $\delta$, as shown in Fig. \ref{scalene_triangle}(b). %
We express the spins $S_i$ on the sublattice $\alpha$ ($\beta, \gamma$ and $\delta$) using the boson operators $\alpha_i$ ($\beta_i, \gamma_i$ and $\delta_i$). %
\begin{table}[t]
\begin{center}
\begin{tabular}{ccccccc}
\hline
\hline
$J_1S$ & $J'_1S$ & $J''_1S$ & $J_2S$ & $J'_2S$ & $J_3S$ & $J'_3S$\\
\hline
$ -0.455$ & $-0.422$ & $-0.150$ & $-0.100$ & $-0.106$ & $-0.338$ & $-0.372$\\
\hline
\hline
$J_zS$ & $J'_zS$ & $DS$ & $ES$ & &\\
\hline
 $-0.187$ & $-0.167$ & $0.160$ & $-0.035$ & &\\
\hline
\end{tabular}
\caption{The Hamiltonian parameters (in meV). }
\label{Table_J}
\end{center}
\end{table}

The procedures of the calculation for the spin-wave energies are essentially the same as those in Ref. \onlinecite{Fishman_CFO_SW}. %
The spin-wave energies at wave vector $q$, $\epsilon_q$, are obtained by solving the Heisenberg equations of motion for the vector $\mbox{\boldmath $v$}_q=(\alpha_q,\beta_q,\gamma_q^\dagger,\delta_q^\dagger,\alpha^\dagger_{-q},\beta_{-q}^\dagger, \gamma_{-q},\delta_{-q})$, where $\alpha_q$, $\beta_q$ and so on are the Fourier-transformed boson operators. %
The equation of motion for $\mbox{\boldmath $v$}_q$ can be written using the $8\times 8$ matrix
$M(q)$ as $id\mbox{\boldmath $v$}_q/dt=-[H,\mbox{\boldmath $v$}_q]=M(q)\mbox{\boldmath $v$}_q$. %
Diagonalizing the matrix $M(q)$, we have obtained the spin-wave energies in the 4SL phase. %

To find a suitable set of magnetic interaction parameters, %
we have prepared $5\times10^8$ sets of the parameters  by Monte Carlo method,\cite{com2} %
and have calculated spin-wave energies $\epsilon_{q_i}$, where $i$ stands for a observed data point. %
For each set of the parameters, we have calculated a sum of squared residuals $\chi^2=\sum_i(E_{q_i}-\epsilon_{q_i})^2$, where $E_{q_i}$ is a observed spin-wave energy, and then have searched a set of the parameters having a small value of $\chi^2$. %
It should be noted that in principle, a finite in-plane anisotropy ($E$ term) splits not only the LE branch but also the HE branch into two branches. %
Therefore, we have four calculated spin-wave branches referred to as LE1-, LE2-, HE1- and HE2-branches. %
Since we could not observed the splitting of the HE-branch in our inelastic neutron scattering measurements, we have used $E_{q_i}$ of the observed HE-branch for both of the calculated HE1- and HE2-branches, in the evaluation of $\chi^2$. %

\begin{figure}[t]
\begin{center}
	\includegraphics[clip,keepaspectratio,width=8cm]{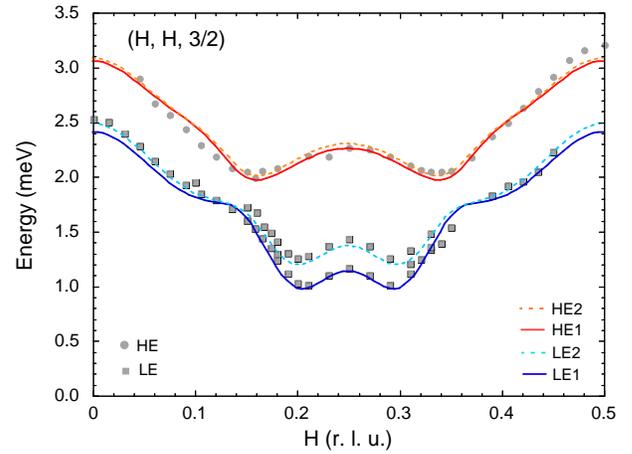}
	\caption{(Color Online) Spin-wave dispersion relations along the $(H,H,\frac{3}{2})$ line calculated with the parameters in Table \ref{Table_J} and the observed spin-wave energies.  }
	\label{disp}
\end{center}
\end{figure}

\begin{figure*}[t]
\begin{center}
	\includegraphics[clip,keepaspectratio,width=.98\textwidth]{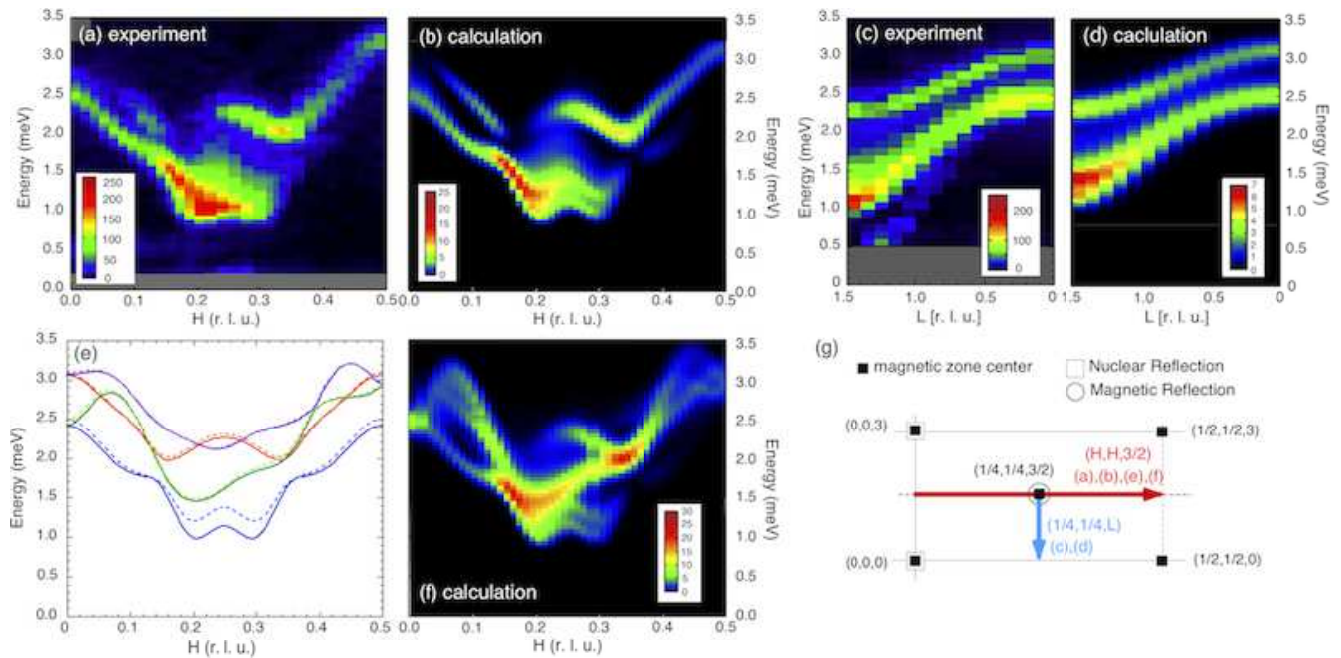}
	\caption{(Color Online) Observed intensity maps of the inelastic neutron scattering measurements along (a) the $(H,H,3/2)$ and (c) $(1/4,1/4,L)$ lines (taken from Ref. \onlinecite{SingleDomainSW}). %
	Calculated intensity maps along (b) the $(H,H,3/2)$ and (d) $(1/4,1/4,L)$ lines for a single domain 4SL phase. %
	(e) Spin-wave dispersion relations for a magnetically multi-domain 4SL phase calculated from the parameters in Table \ref{Table_J}. %
	(f) Calculated intensity map for the multi-domain state in which volume fractions of the three domains are assumed to be equal to each other. %
	(g) The reciprocal lattice map of the $(H,H,L)$ scattering plane. Arrows denote the directions of the $(H,H,3/2)$ and $(1/4,1/4,L)$ lines.}
	\label{convo}
\end{center}
\end{figure*}

Examining the $5\times10^8$ sets of the parameters, we have obtained a set of parameters which successfully reproduces the observed dispersion relations. %
Table \ref{Table_J} shows the values of the Hamiltonian parameters.\cite{com1} %
In Fig. \ref{disp}, we show the calculated spin-wave dispersion relations along $(H,H,3/2)$ line with the observed spin-wave energies.\cite{com3}  %
Using these parameters, we have also calculated inelastic neutron scattering cross-sections described as
\begin{eqnarray}
\frac{d^{2}\sigma}{d\Omega dE'}&\propto&f^2(\bm{\kappa})(1+\hat{\kappa_{z}}^{2})\frac{k_f}{k_i}\sum_{\bm{R},\bm{R'}}e^{-i\bm{\kappa}\cdot(\bm{R}-\bm{R'})}\notag\\
&\times&\int_{-\infty}^{\infty}dt e^{-i\omega t}(\langle S^{+}_{\bm{R}}(0) S^{-}_{\bm{R'}}(t)  \rangle+\langle S^{-}_{\bm{R}}(0) S^{+}_{\bm{R'}}(t)  \rangle)\nonumber\\
\end{eqnarray}
where, ${\bm \kappa}$ is the scattering vector defined as ${\bm \kappa}={\bm k}_f-{\bm k}_i$. %
${\bm k}_f$ and ${\bm k}_i$ are wave vectors of the scattered and incident neutrons, respectively. %
$f({\bm \kappa})$ is the magnetic form factor of a Fe$^{3+}$ ion.\cite{Fe_FormFactor} %
$\kappa_z$ is the $z$-axis component of an unit vector of the scattering vector. %
${\bm R}$ denotes a position of a spin of a Fe$^{3+}$ ions. %
To obtain resolution-convoluted inelastic neutron scattering spectrums, we have employed the Cooper-Nathans type resolution function.\cite{Cooper:a05676} %
As shown in Figs. \ref{convo}(a)-\ref{convo}(d), we have demonstrated that  the calculated intensity maps  show good agreement with the observed data (taken from Ref. \onlinecite{SingleDomainSW}). %

It should be noted that the spin-wave dispersion relations in the $(H,H,L)$ plane does not depend on $J'_2$, because the direction of $J'_2$ is perpendicular to the $(H,H,L)$ plane. %
To determine $J'_2$, we have calculated the spin-wave spectrums in the magnetically multi-domain state. %
Because the directions of $J'_2$ in magnetic domains with the wave vectors of $(-\frac{1}{2},\frac{1}{4},\frac{3}{2})$ and $(\frac{1}{4},-\frac{1}{2},\frac{3}{2})$ are not perpendicular to the $(H,H,L)$ scattering plane, the spin-wave spectrums belonging to these domains contain information on $J'_2$. %
We have adjusted the value of $J'_2$ so that the calculated multi-domain spectrum is close to the observed data in Ref. \onlinecite{Ye_CFO_SW}. %
As a result, $J'_2S$ is determined to be $-0.106$ meV. %
In Figs \ref{convo}(e) and \ref{convo}(f), we show the calculated spin-wave dispersion relations %
and a intensity map for an inelastic neutron scattering measurement in the multi-domain state in which the volume fractions of the three domains are assumed to be equal to each other, respectively. %

We now discuss the determined values of the Hamiltonian parameters. %
Among the nearest neighbor exchange interactions, the magnitude of the exchange interactions connecting two up (or down) spins, $|J''_1|$, is determined to be smaller than $|J_1|$ and $|J'_1|$ which connect up and down spins. 
This is consistent with the nature of the scalene triangle distortion; specifically, the nearest neighbor exchange interactions vary so as to lower the total exchange energy of the 4SL magnetic structure. %
It is worth mentioning that previous x-ray diffraction studies on CFO\cite{Terada_14.5T} and CuFe$_{1-x}$Al$_x$O$_2$\cite{CFAO_Xray} have suggested that this scalene triangle distortion occurs not only in the 4SL phase, but also in the screw-type helimagnetic (multiferroic) phase. %
Haraldsen \textit{et al.} have recently presented an inelastic neutron scattering and theoretical study on magnetic excitations in the helimagnetic phase of Ga-doped CFO.\cite{Haraldsen_INS} %
They have assumed that the oxygen displacements also break the equilateral symmetry of $J_1$ so as to lower the total exchange energy, in the helimagnetic phase. Their calculations have successfully reproduced magnetic excitations from the noncollinear magnetic structure. %
This ensures that the oxygen displacements significantly affect the competing magnetic interactions in this system. %

We have also found that the second and third neighbor in-plane exchange interactions and inter-plane interactions have finite anisotropy, specifically, $J'_2/J_2= 1.05$, $J'_3/J_3=1.10$ and $J'_z/J_z=0.89$. %
This suggests that the effects of the monoclinic structural transition, which are not taken into account in the previous studies,\cite{Ye_CFO_SW,Fishman_CFO_SW,Fishman_stacking,Kimura_ESR_JLTP} are important to explain the spin-wave excitation in the 4SL phase. %
\subsection{Microscopic Picture of the "Electromagnon" Excitation}

As mentioned in introduction, Seki \textit{et al.} have recently discovered the "electromagnon" excitation in the 4SL phase.\cite{Seki_Electromagnon} %
Specifically, their optical spectroscopy have detected two groups of signals corresponding to the zone-center spin-wave modes belonging to the LE- and HE-branches. %
Hereafter, we refer to these spin-wave modes as LE- and HE-modes, respectively. %
Measuring these signals with various light polarization configurations, they have concluded that the HE-mode is "electromagnon" driven by ac electric field, and the LE-mode is conventional "magnon". %
To investigate the mechanism of the "electromagnon" excitation, it is indispensable to establish physical pictures of these spin-wave modes. %
In the following discussion, we have neglected the in-plane anisotropy, $E$, for simplicity, and therefore there are only two spin-wave modes corresponding to the LE- and HE-modes, because the LE1- and LE2-modes (the HE1- and HE2-modes) degenerate without the $E$ term. %

\begin{figure}[t]
\begin{center}
	\includegraphics[clip,keepaspectratio,width=8cm]{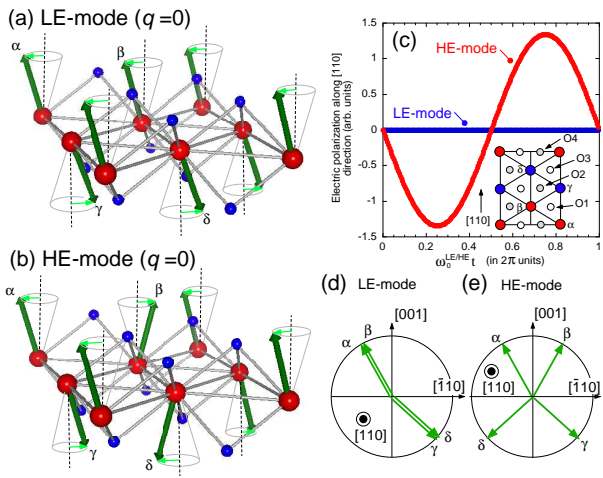}
	\caption{(Color Online) Schematic drawing of the motion of the spins in (a) the LE- and (b) HE-modes. %
	(c) Calculated electric polarization in a magnetic unit cell for the LE- and HE-modes. Inset shows positions of the four inequivalent oxygen sites in a magnetic unit cell. %
	[110]-plane-projections of the spins of the (d) LE- and (e) HE-modes. }
	\label{modes}
\end{center}
\end{figure}

In Figs. \ref{modes}(a) and \ref{modes}(b), we have shown the schematic drawings of the LE- and HE-modes at zone center, which are obtained by the spin-wave analysis without the $E$ term. %
These pictures are consistent with the spin-wave modes presented in the recent ESR study by Kimura \textit{et al}.\cite{Kimura_unpublished} %
Each spin rotates about the $z$ axis with angular frequency of $\omega_0^{\rm LE/HE}$, where $\omega_0^{\rm LE/HE}$ is the spin-wave frequency of the LE/HE-branch at $q=0$. %
In the LE-mode, the up spins (or the down spins) rotate keeping a "collinear" configuration. %
On the other hand, in the HE-mode, the two up spins (or two down spins) are coupled in a "noncollinear" configuration, in a magnetic unit cell. %
In Figs. \ref{modes}(d) and \ref{modes}(e), we show [110]-plane-projections of the spins depicted in Figs. \ref{modes}(a) and \ref{modes}(b), respectively. %
We found that the HE-mode has a screw-like spin texture but the LE-mode does not, as was predicted in Ref. \onlinecite{Seki_Electromagnon}. %
Moreover, the spin-helicity of the screw-like spin texture is reversed by rotating the spins about the $z$ axis by $\pi$. %
Keeping the relationship between the ferroelectricity and the screw-type magnetic ordering in the multiferroic phase of CFO\cite{Kimura_CuFeO2,Kanetsuki_JPCM,Seki_PRB_2007,Ga-induce,CFRO,SpinNoncollinearlity,CompHelicity,CFAO_Helicity} in mind, we anticipate that the noncollinear nature of the HE-mode accounts for the dynamical ME-coupling in the 4SL phase. %

To quantitatively investigate the difference between the LE- and HE-modes, we have applied, to the two modes, the $d$-$p$ hybridization model,\cite{Arima_Symmetry} which successfully explains the ferroelectricity in the screw-type helimagnetic phase of this system. %
This model requires three noncollinear spins surrounding an oxygen ion as a minimal unit to produce uniform electric polarization. %
Hence, we have calculated a sum of electric polarizations at four inequivalent oxygen sites, each of which is surrounded by three Fe ions, in a magnetic unit cell (see inset of Fig.\ref{modes}(c)). %
The procedures of the calculation are essentially the same as those in Refs. \onlinecite{CFAO_Xray} and \onlinecite{Arima_Symmetry}. %
Fig. \ref{modes}(c) shows the calculated electric polarizations along the [110] direction as functions of $\omega_0^{\rm LE/HE}t$ for the LE- and HE-mode, %
revealing that the HE-mode generates electric polarization oscillating with the same angular frequency as $\omega_0^{\rm HE}$, but the LE-mode does not. %
This oscillating electric polarization can be coupled with the ac electric fields. %
We have thus concluded that this must be the microscopic mechanism of the electromagnon excitation in this system. %

\section{Conclusion}

We have investigated spin-wave excitations in the 4SL magnetic ground state of a frustrated magnet CuFeO$_2$. %
By high-energy-resolution neutron inelastic scattering measurements, we have revealed the fine structure of the spin-wave dispersion relations in the 4SL phase. %
Taking account of the trigonal to monoclinic structural transition and the scalene triangle distortion in the 4SL phase, we have developed a model Hamiltonian, and have determined the Hamiltonian parameters. %
We have successfully reproduced the spin-wave dispersion relations and the intensity maps for the inelastic neutron scattering measurements in the single-domain and multi-domain 4SL phase. %
Consequently, we have concluded that the the spin-driven lattice distortions significantly affect the exchange interactions in this system.

Using the results of the spin-wave analysis, we have presented the physical pictures of the zone-center spin-wave modes belonging to the LE- and HE-branches, and have discussed the mechanism of the recently discovered electromagnon excitation. %
The pictures of the LE- and HE-modes suggest that the symmetry of the spin-wave modes are the key to understand this kind of dynamical ME coupling. %
Applying the $d$-$p$ hybridization model\cite{Arima_Symmetry} to the  two modes, we have demonstrated that only the HE-mode generates the oscillating electric polarization, which can be coupled with ac electric field, but the LE-mode does not. %
This indicates that the $d$-$p$ hybridization model, in which spins and electric dipole moments are mediated by a spin-orbit coupling, can explain not only the static coupling between the magnetic ordering and the ferroelectricity, but also dynamical ME-coupling, i.e., electromagnon excitation. %

\section*{Acknowledgments}
This work was supported by a Grant-in-Aid for Young Scientist (B) (Grant No. 23740277), from JSPS, Japan. %
The images of the crystal and magnetic structures in this paper were depicted using the software VESTA developed by K. Momma.\cite{VESTA}

\bibliography{main}

\begin{thebibliography}{33}
\bibitem[{com({\natexlab{a}})}]{com2}
\bibinfo{note}{In the present analysis, we have assumed that $0> J_iS, J'_iS,
  J''_1S >-0.6$, $0<DS<0.2$ and $0>ES>-0.05$ in meV units, where $i=1,2,3,z$.
  The parameter search by Monte Carlo method has been carried out in the above
  parameter regions.}
\bibitem[{com({\natexlab{b}})}]{com1}
\bibinfo{note}{We could not perform a least-square analysis because of strong
  correlations between the Hamiltonian parameters. Therefore, we could not
  estimate the error bars of the parameters shown in Table \ref{Table_J}.}
\bibitem[{com({\natexlab{c}})}]{com3}
\bibinfo{note}{In Ref. \onlinecite{SingleDomainSW}, we have pointed out that
  the spin-wave energies at $(1/2,1/2,0)$ and $(1/4,1/4,3/2)$ are different
  from each other, though both of the two reciprocal lattice points are
  magnetic zone center. At this moment, the reason for the difference between
  the spin-wave energies at $(1/2,1/2,0)$ and $(1/4,1/4,3/2)$ is not clear. In
  the present analysis, we have determined the Hamiltonian parameters so that
  the calculated and observed spin-wave energies agree with each other at
  $(1/4,1/4,3/2)$.}
\end{thebibliography}

\end{document}